\documentclass[runningheads]{llncs}
\usepackage[T1]{fontenc}
\usepackage{graphicx}
\usepackage{array}
\usepackage{booktabs}
\usepackage{threeparttable}
\usepackage{multicol}
\usepackage{hyperref}

\usepackage{xcolor}
\definecolor{codegreen}{rgb}{0,0.6,0}
\definecolor{codegray}{rgb}{0.5,0.5,0.5}
\definecolor{framegray}{rgb}{0.8,0.8,0.8}
\definecolor{codepurple}{rgb}{0.58,0,0.82}
\definecolor{backcolour}{rgb}{0.95,0.95,0.92}

\usepackage{listings}
\lstdefinestyle{mysparql}{
  language=SQL,
  morecomment=[l]{\#},
  morekeywords={OPTIONAL, FILTER, VALUES, SERVICE, OFFSET},
  moredelim=*[s][\color{codegray}]{wdt:}{\ },
  moredelim=*[s][\color{codegray}]{wd:}{\ },
  commentstyle=\color{codegreen},
  keywordstyle=\color{codepurple},
  numberstyle=\tiny\color{codegray},
  stringstyle=\color{magenta},
  basicstyle=\ttfamily\footnotesize,
  frame=single,
  rulecolor=\color{framegray}
}
\newcommand{\wdata}[1]{{\small\texttt{#1}}}

\begin{document}
\title{Categorizing Mathematical Concepts with LLM Voting Ensembles in Mathswitch}
\titlerunning{Categorizing Mathematical Concepts}
\author{Katja Berčič\inst{1, 2} \and
Slobodan Stanojevikj\inst{1}}
\authorrunning{K. Berčič and S. Stanojevikj}
%
\institute{Faculty of Mathematics and Physics, University of Ljubljana, Ljubljana, Slovenia
\and Institute of Mathematics, Physics and Mechanics, Ljubljana, Slovenia}
\maketitle              
\begin{abstract}
Mathswitch is an open-source project that imports mathematical concept records from sources such as 
Wikidata, Wikipedia, MathWorld, Encyclopedia of Mathematics, nLab, ProofWiki, and Agda-Unimath, 
and links records that refer to the same concept. It does not reorganize or redefine the imported content; 
each source retains its own structure. 
The current focus is on importing high-quality concept data from Wikidata and the resources it links to, 
with plans to expand to further sources and better concept linking. 
Because the concept set is approximated through queries over Wikidata's collaboratively edited graph, 
the imported data is noisy: some items are non-mathematical, while others are ambiguous. 
In this paper, we test whether a voting ensemble of LLM judges can filter this noise. 
We evaluate it on Wikidata items with known MathWorld identifiers as a positive control, 
and examine how classification changes when database identifiers are removed from context. 
We then inspect the cases where the judges disagree with MathWorld and group these disagreements into three categories (degenerate descriptions, narrow scope bias, and editorial-scope mismatches) that suggest different remediation strategies.

\keywords{Mathswitch \and categorization \and LLM-as-judge \and Wikidata \and SPARQL \and mathematical knowledge bases.}

\end{abstract}
\section{Introduction}

Mathematical concepts can be found in many kinds of resources. Encyclopedias such as MathWorld and the Encyclopedia of Mathematics provide curated reference entries. Collaborative platforms such as Wikipedia, nLab, and ProofWiki offer community-maintained expositions. Wikidata provides a structured knowledge graph, the LMFDB documents its objects through knowls, and libraries of formalized mathematics such as Mathlib, Mizar, and Agda-Unimath contain machine-checked definitions and proofs. Each resource uses its own vocabulary, level of formality, and organizational scheme, and a concept described in one source cannot be systematically located in another.

Mathswitch is an open-source project that works toward connecting mathematical concepts across such resources. It imports concept records from several sources and groups records that appear to refer to the same idea. It does not reorganize or redefine the imported content; each source retains its own structure, and Mathswitch makes the connections between them explicit. The current focus is on importing high-quality concept data from Wikidata and the resources it links to, but the set of covered sources and the sophistication of concept linking are both intended to grow.

Cross-referencing at this scale depends on the quality of the imported data. There is no authoritative registry of all mathematical concepts, so Mathswitch approximates the concept set through hand-tuned queries over Wikidata's collaboratively edited graph. The resulting data is noisy: some items are clearly non-mathematical but match the query patterns, while others are genuinely ambiguous terms that carry both mathematical and non-mathematical meanings. This noise degrades the reliability of the cross-reference network.

\begin{figure}[ht!]
  \centering
  \includegraphics[width=0.8\textwidth]{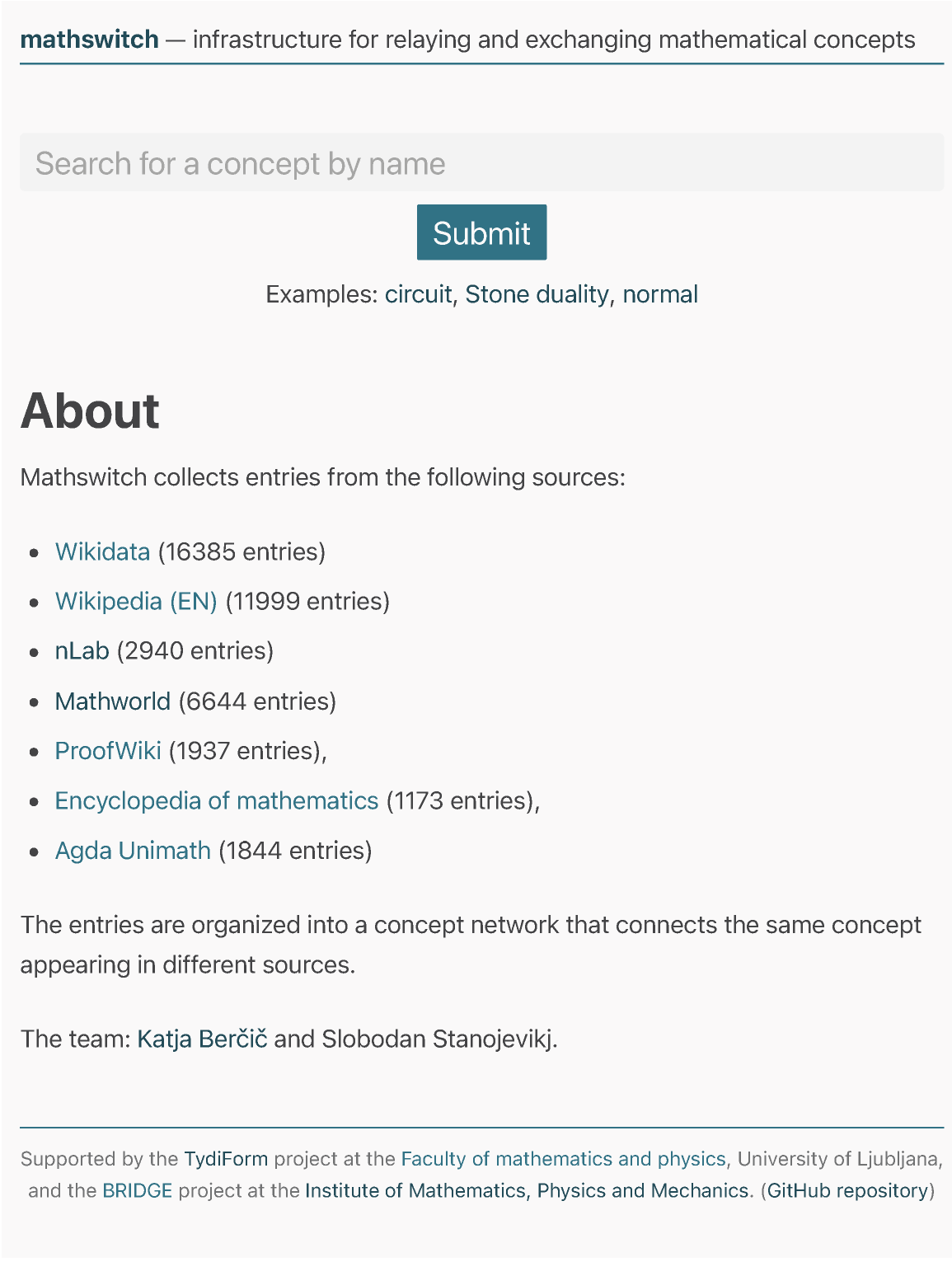}
  \caption{The Mathswitch landing page with a search box and a list of sources together with the number of concepts imported from each.}
  \label{fig:landing}
\end{figure}

\begin{figure}[ht!]
  \centering
  \includegraphics[width=0.8\textwidth]{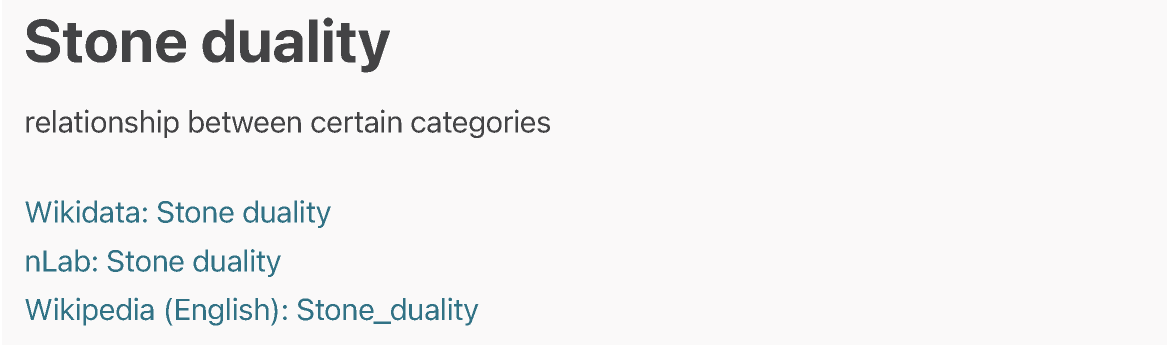}
  \caption{The Mathswitch concept page for Stone duality.}
  \label{fig:concept}
\end{figure}

Mathswitch is an open-source web application, available at \url{mathswitch.xyz}, that serves pages about mathematical concepts, linking records across external mathematical resources. The landing page offers a search interface over the concept set (Figure~\ref{fig:landing}), and each concept page collects links to Wikipedia, nLab, MathWorld, ProofWiki, the Encyclopedia of Mathematics, and the formal mathematics library Agda-Unimath for each concept (when these links are available), together with the Wikidata label and description (Figure~\ref{fig:concept}). The current prototype is built around Wikidata and the identifiers it exposes.

In this paper, we describe the Mathswitch data pipeline and evaluate an automated categorization approach based on a voting ensemble of LLM judges. The pipeline consists of three stages:
\begin{enumerate}
  \item preparing the SPARQL queries (Section~\ref{sec:sparql}),
  \item fetching data from Wikidata and Agda-Unimath (Section~\ref{sec:fetching-concepts}),
  \item filtering concepts with the voting ensemble and building the concept network (Section~\ref{sec:filtering-concepts}).
\end{enumerate}

We evaluate the ensemble on Wikidata items with known MathWorld identifiers as a positive control and examine how classification changes when database identifiers are removed from context. As a negative control, we run the same ensemble on a sample of physics concepts from Wikidata. We then inspect the cases where the judges disagree with MathWorld and group them into three categories (Section~\ref{sec:qualitative}).

\section{Related work}

Several systems organize and connect mathematical knowledge across sources. Some build unified formal representations: OMDoc provides a markup language for mathematical documents \cite{kohlhase2006omdoc}, the MMT framework generalizes this with a foundation-independent module system \cite{rabe2013scalable}, MathHub uses MMT for hosting and cross-referencing libraries \cite{iancu2014mathhub}, and the Formal Abstracts project proposes stating theorems in a controlled formal language \cite{FABstracts:on}. Gauthier and Kaliszyk \cite{gauthier2019aligning} and M\"uller et al.\ \cite{muller2017classification} align concepts across proof assistant libraries without a common formalism.

Wikidata is collaboratively edited, and its quality varies across topics \cite{piscopo2019Wikidata,shenoy2022Wikidata}. Instance-of and subclass-of assignments are often incomplete or inconsistent, which affects downstream queries. For mathematical content, the MaRDI initiative builds a Wikibase-powered knowledge graph from sources such as DLMF, zbMATH, and arXiv \cite{schubotz2023bravo}. Scharpf, Schubotz, and Gipp link mathematical formulae to Wikidata entities through the AnnoMathTeX system \cite{scharpf2018representing,scharpf2021fcd}. The Wikidata WikiProject Mathematics maintains curated property lists and modeling conventions \cite{Wikidata_wikiproject_math}. The OntoMath\hbox{-}PRO ontology provides a curated vocabulary for mathematical concepts \cite{nevzorova2014ontomathpro}. Work on Wikidata subsetting \cite{beghaeiraveri2023subsetting,nguyen2022Wikidatalite,Wikidata_subsetting} shows that strict domain-based extraction is hard, since concepts do not separate cleanly into domains. Mathswitch faces a narrower version of this problem: given a candidate set from SPARQL queries, it needs to decide which items are mathematical concepts.

Recent work uses large language models as automated evaluators, asking them to judge whether a given output satisfies a quality criterion or matches a reference answer. Zheng et al.\ show that strong language models approximate human preference judgments and document systematic biases (position, verbosity, self-preference) in single-model judges \cite{zheng2023judging}. Sampling multiple responses from a single model and majority-voting over them improves reliability \cite{wang2023selfconsistency}, but does not address biases shared by all samples from the same model. Querying several different models goes further: LLM-Blender ranks and fuses outputs from a model pool \cite{jiang2023llmblender} and Li et al.\ report accuracy gains from majority voting across multiple agents \cite{li2024moreagents}. Most of this work targets open-ended generation and preference ranking.

\section{Wikidata, SPARQL queries and their results}
\label{sec:sparql}

Wikidata is a collaboratively edited knowledge graph. Its main building blocks are \emph{items}, \emph{properties}, and \emph{lexemes}. Items represent entities and are identified by Q identifiers: for example, \wdata{Q11348} is the mathematical concept \emph{function} and \wdata{Q395} is \emph{mathematics} (as a field of study). Properties represent relations and are identified by P identifiers: \wdata{P31} is \emph{instance of}, \wdata{P2579} is \emph{studied by}, and \wdata{P2812} is \emph{MathWorld identifier}. Lexemes (L identifiers) represent words and phrases; Mathswitch does not use them. A statement (subject--predicate--object) links an item to a value through a property. For instance, the statement \wdata{Q11348}~\wdata{P2579}~\wdata{Q395} means that \emph{function}(s) are studied by \emph{mathematics}. External identifiers such as \wdata{P2812} link Wikidata items to records in other databases.

Mathswitch queries this knowledge graph using SPARQL to form the backbone of the set of mathematical concepts. Two queries are built around different property paths. Each query also asks for the external identifiers listed in Table~\ref{tab:ext-ids}, when they are present on an item. A third pass fetches any Wikidata item that carries one of these identifiers, as items with known external links are likely relevant even if they do not match any of the property paths. The listing in Appendix~\ref{app:sparql} shows a full example query for one of the property paths.

Because the queries inevitably return some non-mathematical items, each query also excludes certain categories of items that were found to cause noise (Table~\ref{tab:exclusions}). These exclusions are maintained manually and extended as new sources of noise are noticed.

\begin{table}[htb]
    \centering
    \caption{External identifiers requested by each SPARQL query. These are fetched when available but are not required for an item to be returned.}
    \label{tab:ext-ids}
    \begin{tabular}{p{0.25\textwidth}p{0.5\textwidth}}
        \toprule
        \textbf{Property} & \textbf{External source} \\
        \midrule
        \wdata{P4215} & nLab \\
        \wdata{P2812} & MathWorld \\
        \wdata{P6781} & ProofWiki \\
        \wdata{P7554} & Encyclopedia of Mathematics \\
        \bottomrule
    \end{tabular}
\end{table}

\begin{table}[htb]
    \centering
    \caption{Excluded Wikidata categories at the time of writing. These are added manually as they are noticed.}
    \label{tab:exclusions}
    \begin{tabular}{p{0.25\textwidth}p{0.5\textwidth}}
        \toprule
        \textbf{Wikidata Item} & \textbf{Excluded category} \\
        \midrule
        \wdata{Q21199} & natural numbers \\
        \wdata{Q28920044} & positive integers \\
        \wdata{Q6256} & countries \\
        \wdata{Q714737} & philosophical concepts \\
        \wdata{Q5} & humans \\
        \bottomrule
    \end{tabular}
\end{table}

The two main Wikidata queries capture different patterns of how mathematical concepts can be linked to mathematical topics and areas. Each follows a short property path through the graph.
\begin{enumerate}
    \item \textbf{Instances of a topic studied by mathematics.} The item is an instance (\wdata{P31}) of some topic, and that topic is studied by (\wdata{P2579}) mathematics (\wdata{Q395}). For example, \emph{infinity} (\wdata{Q205}) is an instance of \emph{mathematical concept} (\wdata{Q24034552}), which is studied by mathematics.
    \begin{lstlisting}
  ?item wdt:P31 ?topic .
  ?topic wdt:P2579 wd:Q395 .
    \end{lstlisting}
    \item \textbf{Concepts studied by a branch of mathematics.} The item is studied by an area (\wdata{P2579})  and that area is an instance of a branch of mathematics (\wdata{Q1936384}). For example, \emph{Euler's formula} (\wdata{Q184871}) is studied by \emph{complex analysis} (\wdata{Q193756}), which is an instance of a branch of mathematics.
    \begin{lstlisting}
  ?item wdt:P2579 ?area .
  ?area wdt:P31 wd:Q1936384 .
    \end{lstlisting}
\end{enumerate}

\paragraph{Notes on Wikidata SPARQL query results.}
The items returned by these queries are noisy, despite the hand-tuned queries and exclusion filters. Items that are not mathematical can match the patterns if they have been incorrectly tagged in Wikidata, or if they are legitimately associated with a mathematical area without themselves being mathematical. Conversely, genuine mathematical concepts whose Wikidata entries lack the relevant properties will be missed entirely. This creates a trade-off between a broader query, with fewer missed concepts, and a more precise one, with fewer non-mathematical items included.

A representative example is \wdata{Q4164871}, whose primary label is \emph{position} (``social role with a set of powers and responsibilities within an organization''), but which also carries \emph{function} as an alias. This alias overlaps with the label of \wdata{Q11348} (the mathematical concept of \emph{function}, described as ``association of a single output to each input''), and some Wikidata editors conflate the two. The consequence is that holders of administrative positions, items that are instances of \wdata{Q4164871}, can end up linked into the mathematical property chains and appear in Mathswitch's import. Filtering out humans via \wdata{FILTER NOT EXISTS \{ ?item wdt:P31 wd:Q5 \}} mitigates this partially, but a person whose entry does not carry \wdata{Q5} directly as an instance type can still pass the filter.

A second, more subtle problem is that certain terms can be legitimately associated with both mathematical and non-mathematical meanings. For example, \emph{tree} appears in graph theory as a well-established mathematical structure, but also exists as a biological concept; similarly, \emph{group} is fundamental in abstract algebra yet is a common word in everyday language and other scientific disciplines. Wikidata metadata alone is often insufficient to decide whether such an item refers to the mathematical sense.

\section{Fetching data}
\label{sec:fetching-concepts}

The prepared SPARQL queries are sent to the Wikidata Query Service endpoint, which returns JSON result bindings. 
For each item that has an associated English Wikipedia article, 
the pipeline fetches the full article text via the Wikipedia API.

The fetched JSON results are converted into \texttt{Item} records through a polymorphic class hierarchy, 
where each source (Wikidata, Wikipedia, nLab, MathWorld, ProofWiki, Encyclopedia of Mathematics) 
has its own subclass responsible for extracting the correct identifier, URL, and name from the raw data. 
When a Wikipedia article text is available, the pipeline applies named entity recognition (NER) 
using spaCy with the model \texttt{en\_core\_sci\_lg}~\cite{neumann-etal-2019-scispacy} to extract domain-relevant keywords, 
which are stored alongside the item. 
Items are stored in the database, with duplicates detected and skipped via integrity constraints. 
Cross-source links are then created between items that originate from the same Wikidata entity.

\paragraph{Agda-Unimath concepts.}
Agda-Unimath data is fetched independently from its public JSON endpoint.
The maintainers of the Agda-Unimath library of formalized mathematics in Agda
manually annotate concepts in their library~\cite{agda_unimath_concepts}, 
and share these annotations with us.
The concepts are annotated directly in the source files using a preprocessor macro as shown below. 
\begin{verbatim}
A {{#concept "circuit" Agda=circuit-Undirected-Graph WD="Cycle" 
    WDID=Q245595}} in an ...
\end{verbatim}
Each annotation records a concept name, an optional link to the Agda definition, 
and an optional Wikidata identifier. 
A preprocessor collects these annotations into a \texttt{concept\_index.json}
\footnote{\url{https://unimath.github.io/agda-unimath/concept_index.json}}
file with fields for the name, Wikidata identifier, definition link, and a stable opaque identifier. 
Mathswitch imports this index to obtain Agda-Unimath concepts and their Wikidata links. 
Because the annotations are added by hand, they are high-quality but limited to concepts that 
contributors have chosen to tag.

\section{Filtering concepts and building the concept network}
\label{sec:filtering-concepts}

No static set of filters can fully solve a problem rooted in the incompleteness and inconsistency of a collaboratively edited knowledge graph. This section describes an automated, per-item categorizer based on a voting ensemble of LLM judges.

Building a dedicated classifier for this task would require labeled training data, a model architecture, and a benchmark to evaluate against. No labeled dataset of ``is this a mathematical concept'' judgments exists, and producing one by hand is not straightforward: the boundary cases that matter most sit at the edges of subfields and can require specialized domain expertise to resolve. An LLM ensemble does not require labeled data. Each model already encodes information about mathematical language, and aggregating votes from several models reduces the impact of any single model's errors. The experiment in this paper is a feasibility test of this approach.

The categorizer sends each imported item to several independent LLMs and records their votes. 
Each judge receives a structured prompt with the item's name, description, keywords, and article text. 
It must answer a yes/no question, \emph{``Is the given concept a mathematical concept?''}, 
with an integer confidence score between 0 and 100. Using more than one model reduces the impact of failure modes in any single one. 
Aggregating the confidence scores gives a threshold that controls the precision/recall trade-off downstream. 
The ensemble is heterogeneous and configurable. 
The categorizer also supports API-based backends such as OpenAI's GPT-4 and Anthropic's Claude.

The judges' votes feed back into the data pipeline. Items classified as non-mathematical by a majority of judges are appended to the SPARQL exclusion list on the next import. Items that the ensemble has already filtered out do not reappear on subsequent runs.

The categorizer is implemented as a Django management command backed by two services: a \texttt{CategorizerService} that orchestrates the process and an \texttt{LLMService} that dispatches calls to individual models.

\paragraph{Prompt construction.}
For each item, the \texttt{CategorizerService} assembles a structured prompt consisting of three parts. First, a system preamble instructs the model to act as a categorization judge. Two preamble variants exist: a simple format that asks for a comma-separated answer and confidence score (e.g.\ \texttt{yes,85}), used with smaller models, and an extended format that asks for three labeled lines (answer, confidence, and explanation), used with larger models where the additional output is affordable. Appendix~\ref{app:prompt} shows the extended variant. Second, the item's metadata is included: its name, the first 100 characters of its description, up to 200 characters of extracted keywords, and up to 1000 characters of article text. These cutoffs keep the input within the context window of the smallest model in the ensemble. Third, the predicate to evaluate is appended: \emph{``Is the given concept a mathematical concept, given the name, description, keywords, and article text?''}

The full prompt template, together with a filled-in instance for the Wikidata item \emph{Wiener sausage}, is given in Appendix~\ref{app:prompt}. This item is chosen only to illustrate the template, and it happens to have one of the degenerate descriptions discussed in Section~\ref{sec:qualitative}. In a typical input, the keywords and article text slots carry more content than they do in the example.

\paragraph{Judge pool and model dispatch.}
The \texttt{LLMService} provides a unified interface over multiple LLM backends. Each backend is registered as a handler keyed by an \texttt{LLMType} enum value. The default judge pool used in this paper is three locally-hosted instruction-tuned models served via Ollama: DeepSeek-R1 14B, Gemma-3 12B, and Qwen-2.5 14B. The service also supports API-based backends, including OpenAI's GPT-4 and Anthropic's Claude, all configurable through environment variables.

\paragraph{Response parsing.}
Each model's raw output is parsed into a boolean answer and an integer confidence score. The parser accepts multiple separators (comma, comma-space, or space). Recognized answer formats are \texttt{yes}/\texttt{no}, \texttt{true}/\texttt{false}, and \texttt{1}/\texttt{0}. If the model fails to produce a parseable confidence value, a default of 50 is assigned. Responses that fall outside the 0--100 range are similarly defaulted.

\paragraph{Persistence and feedback loop.}
Each judge's result is stored as a record in the database (\texttt{CategorizerResult}), linked to the evaluated item. The record stores the LLM type, raw response text, parsed boolean answer, and confidence score. If one judge fails (e.g., due to a model loading error or timeout), the remaining judges still proceed, ensuring partial results are captured.

On the next data import, the pipeline queries the \texttt{CategorizerResult} table for all Wikidata items where a majority of judges answered \texttt{no}. These item identifiers are appended to the SPARQL exclusion list alongside the hardcoded exclusions, so that items previously judged as non-mathematical are automatically filtered out of future queries. This creates a feedback loop: each categorization run refines the data loaded in subsequent imports.

\paragraph{Building the concept network.}
Finally, the system groups items into unified \texttt{Concept}\emph{s} using a Union-Find algorithm. All items and their cross-source links are loaded into the Union-Find data structure, which computes connected components. Each connected component (a set of items from different sources that refer to the same mathematical idea) becomes a single Concept. Items with no links to other sources form singleton Concepts. An additional \texttt{link\_same} step can create links between items that share the same name across sources, which are then merged in a subsequent clustering pass.

\section{Results}
\label{sec:results}

The Wikidata import (April 2026) returned 16\,385 items from the two SPARQL queries. Of these, 6644 carry a MathWorld identifier and 9741 do not. We drew two proportional random samples: 1000 items from the MathWorld-backed pool and 1500 items from the pool without a MathWorld identifier, sampled at the same rate. The MathWorld-backed sample serves as a positive control, since items curated by MathWorld editors are expected to be mathematical. The second sample has no ground truth: it is used to measure the rate at which the judge pool flags untagged items as non-mathematical and to surface candidates for the SPARQL exclusion list.

Table~\ref{tab:per-judge} reports per-judge classification rates on the MathWorld-backed sample. The \emph{Included} row uses the prompt as described in Section~\ref{sec:filtering-concepts}: all Wikidata metadata is passed to the judges, including database identifiers. The \emph{Excluded} row is the same experiment with any reference to MathWorld removed from the item data before prompt construction, to test whether the judges rely on surface markers of database membership. The drop is about one percentage point per judge, which suggests the judges are not using the MathWorld string as a shortcut.

\begin{table}[ht!]
    \begin{threeparttable}
        \caption{Percentage of items in the 1000-item MathWorld-backed sample classified as mathematical by each judge. \emph{Included}: full item metadata. \emph{Excluded}: surface references to MathWorld are removed from the prompt context before evaluation.}
        \label{tab:per-judge}
        \begin{tabular}{p{0.24\textwidth}>{\raggedleft\arraybackslash}p{0.24\textwidth}>{\raggedleft\arraybackslash}p{0.24\textwidth}>{\raggedleft\arraybackslash}p{0.24\textwidth}}
            \toprule
            MathWorld IDs & Ollama DeepSeek\textsuperscript{a} & Ollama Gemma\textsuperscript{b} & Ollama Qwen\textsuperscript{c} \\
            \midrule
            Included & 99.4\% & 98.2\% & 99.1\% \\
            Excluded & 98.3\% & 96.0\% & 98.0\% \\
            \bottomrule
        \end{tabular}
        \begin{tablenotes}
        \footnotesize
        \item[a] DeepSeek-R1 14B
        \item[b] Gemma-3 12B
        \item[c] Qwen-2.5 14B  
        \end{tablenotes}
    \end{threeparttable}
\end{table}

Table~\ref{tab:confusion} breaks down both samples by majority ensemble vote against the MathWorld-backed label. On the 1000-item positive control, 982 items are classified as mathematical and 18 are not, giving an apparent ensemble accuracy of $98.2\%$ against the MathWorld-backed label. The \emph{No MathWorld ID} row should not be read as an error count: the 795 items classified as mathematical are candidates the ensemble endorses, and the 705 items classified as non-mathematical are candidates for the SPARQL exclusion list. The 795/1500 rate is a flagging rate on untagged items, not a false-positive rate.

\begin{table}[ht!]
    \centering
    \caption{Majority ensemble vote on both samples. The \emph{Yes} row is the 1000-item MathWorld-backed positive control. The \emph{No} row is the 1500-item proportional sample of items returned by the SPARQL queries without a MathWorld identifier, for which no ground truth is available (and should not be interpreted as false positives and true negatives).}\label{tab:confusion}
    \begin{tabular}{p{0.18\textwidth}>{\raggedleft\arraybackslash}p{0.18\textwidth}>{\raggedleft\arraybackslash}p{0.18\textwidth}}
        \toprule
        & \multicolumn{2}{l}{Classified as mathematical} \\
        \midrule
        MathWorld ID & Yes & No \\
        \midrule
        Yes & 982 &  18 \\
        No  & 795 & 705 \\
        \bottomrule
    \end{tabular}
\end{table}

\subsection{ROC analysis}

The per-judge and majority-vote results above treat classification as a binary decision. To evaluate how well the ensemble separates mathematical from non-mathematical items across all possible thresholds, we compute a ROC curve over a continuous score derived from the judges' confidence values.

For each judgment, the raw confidence is converted to a math score: if the judge answered \emph{yes}, the math score equals the confidence; if the judge answered \emph{no}, the math score is $100 - \mathit{confidence}$. A confident \emph{no} (e.g.\ confidence 90) thus becomes a math score of 10. The per-item score is the mean of the math scores across the three judges, divided by 100 to obtain a value in $[0, 1]$. This score is passed to \texttt{sklearn.metrics.roc\_curve} as the predicted probability, with the ground-truth label being whether the item carries a MathWorld identifier.

\begin{figure}[ht]
  \centering
  \includegraphics[width=0.8\textwidth]{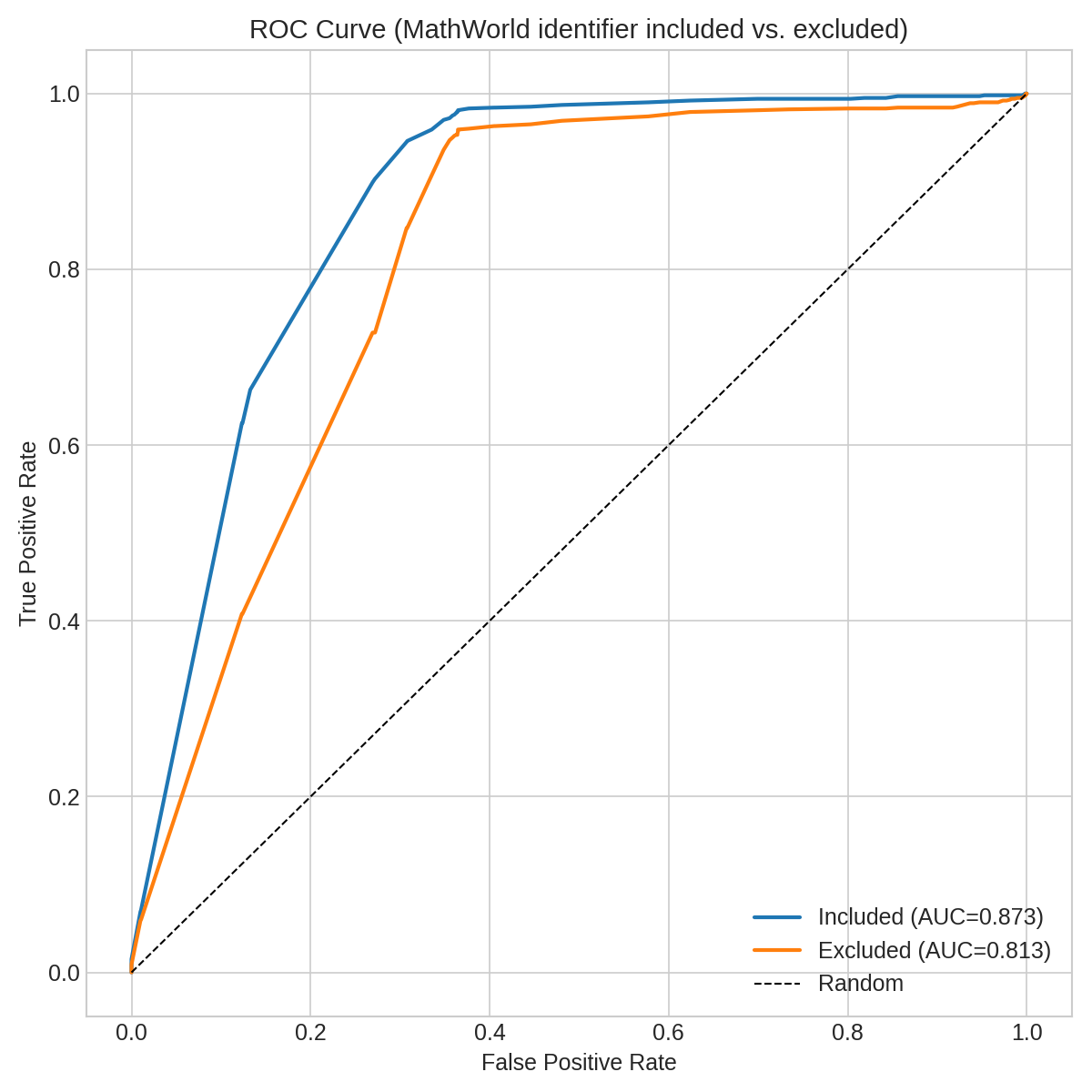}
  \caption{ROC curves for the ensemble math score on the combined 2500-item sample. \emph{Included}: full item metadata. \emph{Excluded}: MathWorld references removed from prompt context.}
  \label{fig:roc}
\end{figure}

Figure~\ref{fig:roc} shows two ROC curves. The first (Included) uses the full item metadata, including any MathWorld identifiers present in the prompt. The second (Excluded) removes surface references to MathWorld before prompt construction. Both curves confirm that the ensemble discriminates well between the two groups. The small gap between the two curves is consistent with the one-percentage-point drop observed in Table~\ref{tab:per-judge}, indicating that the judges do not rely on the MathWorld string as a classification shortcut.

\subsection{Control experiment on physics concepts}

As a negative control, we ran an analogous SPARQL query for physics concepts on Wikidata and submitted a sample of 500 items to the same ensemble. Table~\ref{tab:physics} shows the distribution of votes. Of the 500 items, 430 (86\%) received a unanimous \emph{no} from all three judges. Only 24 items (4.8\%) were classified as mathematical by a majority of judges. The 10 items on which all three judges agreed are boundary cases where physics and mathematics overlap (e.g.\ differential equations, symmetry groups). This confirms that the ensemble does not indiscriminately accept items that are adjacent to mathematics.

\begin{table}[ht]
    \centering
    \caption{Distribution of judge votes classifying items as mathematical or not, on a 500-item sample of Wikidata physics concepts.}\label{tab:physics}
    \begin{tabular}{>{\raggedleft\arraybackslash}p{0.28\textwidth}>{\raggedleft\arraybackslash}p{0.20\textwidth}>{\raggedleft\arraybackslash}p{0.20\textwidth}}
    \toprule
    Judges voting ``math'' & Number of items & Mean confidence \\
    \midrule
    0 & 430 &  86.3 \\
    1 &  46 &  80.5 \\
    2 &  14 &  82.4 \\
    3 &  10 &  89.0 \\
    \bottomrule
\end{tabular}

\end{table}

The mean confidence column refines this picture. Judges are most certain when they agree: the unanimous \emph{no} group (86.3) and the unanimous \emph{yes} group (89.0) carry the highest confidences, while the split groups (80.5 and 82.4) sit noticeably lower, reflecting genuine uncertainty on items the ensemble could not cleanly sort. The 10 items in the unanimous-\emph{math} row are listed in Table~\ref{tab:physics-unanimous}. Rather than misclassifications, they are concepts that live on the physics--mathematics boundary: nonlinear PDEs (Davey--Stewartson, Haselgrove, Jeans equations), differential-geometric structures (Fisher--Kähler manifold), operator-theoretic quantities (Hankel singular value), dynamical systems (Kepler orbit, Lamb--Chaplygin dipole), and an information-theoretic invariant (entropy rate).
That the ensemble's highest-confidence ``false positives'' are precisely the items a mathematician would also flag as mathematical is further evidence that the judges are tracking mathematical content rather than surface cues.

\begin{table}
    \centering
    \caption{The ten Wikidata physics concepts on which all three judges unanimously voted ``math'', with mean confidence averaged across the three judgments. These items lie on the physics--mathematics boundary (differential equations, manifolds, operator theory, dynamical systems).}\label{tab:physics-unanimous} 

\begin{tabular}{>{\raggedright\arraybackslash}p{0.50\textwidth}>{\raggedleft\arraybackslash}p{0.20\textwidth}}
    \toprule
    Concept & Mean confidence \\
    \midrule
    Airy spheroid &  88.3 \\
    Bubble of nothing &  86.7 \\
    Davey–Stewartson equation &  93.3 \\
    Entropy rate &  88.3 \\
    Fisher–Kähler manifold &  91.7 \\
    Hankel singular value &  93.3 \\
    Haselgrove's equations &  83.3 \\
    Jeans equations &  86.7 \\
    Kepler orbit &  86.7 \\
    Lamb–Chaplygin dipole &  91.7 \\
    \bottomrule
\end{tabular}

\end{table}

\subsection{Qualitative analysis of disagreements on MathWorld items}
\label{sec:qualitative}

To understand where the ensemble disagrees with the MathWorld-identifier positive control, we inspected the 48 Wikidata items that carry a MathWorld identifier and for which at least one judge returned \emph{no}. Appendix~\ref{app:example-disagreements} shows selected examples of judge responses. Three distinct phenomena emerge, with different implications for the pipeline.

\paragraph{One model drives most of the dissent.}
Gemma votes \emph{no} on 39 of the 48 items (81\%), compared with 20 for Qwen (42\%) and 17 for DeepSeek (35\%). In most cases Gemma is the sole minority vote against the other two, which matches its slightly lower rate in Table~\ref{tab:per-judge} and shows that the aggregate per-judge accuracies conceal a single model driving most of the disagreement. The confusion matrix in Table~\ref{tab:confusion} reflects this only in part, since majority voting absorbs many of the Gemma-only dissents before they become ensemble errors.

\paragraph{Degenerate Wikidata descriptions.}
Twelve of the 48 items have descriptions that carry no domain content: nine are the literal string \emph{mathematical concept} (for example 
\emph{Wiener sausage}, \wdata{Q7999139}; 
\emph{principle of permanence}, \wdata{Q7245167}; 
\emph{fence}, \wdata{Q5442977}),
and three are empty (e.g.\
\emph{Landau's constants}, \wdata{Q2994932}).
For these items the judges effectively receive only the bare name and have no domain anchor to work from. Gemma treats the tautological hint as low evidence and overrides it from surface-name associations at confidence $\geq 90$: \emph{Wiener sausage} is classified as food, \emph{fence} as civil engineering, \emph{principle of permanence} as Piaget's object permanence, and \emph{infinite loop space machine} (\wdata{Q16926157}) as science fiction. These misclassifications follow from the missing context. The underlying cause is a data-quality problem upstream of the categorizer.

\paragraph{Gemma rejects items with applications outside of mathematics.}
On items whose descriptions do mention a physical, biological, or computational context, Gemma applies a consistent pattern, visible in the explanation field of its responses: if the description names an applied field, the concept is treated as an \emph{application} of mathematics rather than mathematics itself, regardless of how canonically mathematical the object is. 
For example, this pattern flags \emph{Pauli matrices} (\wdata{Q336233}, ``matrices important in quantum mechanics''),
\emph{Fisher's equation} (\wdata{Q1763840}, ``PDE named after statistician and biologist Ronald Fisher''), and
\emph{margin of error} (\wdata{Q1352827}).
DeepSeek and Qwen weigh the name and the description together and typically side with the mathematical reading, with a small number of Qwen exceptions on physics-flavored items.

\paragraph{Legitimate editorial-scope mismatches.}
The ten items on which all three judges unanimously answered \emph{no} (e.g.\
\emph{stochastic resonance}, \wdata{Q1999781}, and
\emph{window function}, \wdata{Q1404885})
fall into three groups: signal-processing tooling, mechanical or materials applications, and historical or recreational entries. They reflect a genuinely wider editorial scope in MathWorld than the LLM judges are willing to attribute to mathematics on the basis of a one-line description. These items are informative about the reference corpus, not about judge quality.

\section{Discussion and future work}

A voting ensemble of three small, locally hosted LLMs correctly classifies 98.2\% of MathWorld-backed items as mathematical, with only a one-percentage-point drop when surface references to MathWorld are removed from the prompt. The ensemble is usable as a filter in the Mathswitch pipeline. The more informative result is where and why it fails.

The three phenomena in Section~\ref{sec:qualitative} are separable and call for different fixes. Degenerate-description items should be filtered or enriched before reaching the judges, since the judges then have only the item name to work from. This accounts for most of Gemma's dissent. The narrow scope bias affects Gemma but not the other two models. Majority voting overrides these cases, so they rarely change the ensemble outcome. The editorial-scope items caution against treating a MathWorld identifier as a silver standard: a small fraction of MathWorld-backed items are edge cases that a narrower working definition of \emph{mathematical concept} would legitimately exclude.

\paragraph{Limitations.}
The evaluation in Section~\ref{sec:results} relies on two assumptions that are worth stating explicitly. First, MathWorld-backed items are treated as positive examples, and the editorial-scope mismatches show that this label is noisy at the boundaries of what is a mathematical concept. Accuracies computed against this label therefore carry a small irreducible error. Second, the 1500-item sample of items without a MathWorld identifier has no ground truth. We report the ensemble's flagging rate on this sample but do not claim it measures accuracy. Finally, the experiment is run on a single Wikidata snapshot from April 2026; stability across snapshots is not measured.

\paragraph{Future work.}
Three follow-ups are on the near-term list. Enriching degenerate Wikidata descriptions from the linked Wikipedia article before prompt construction should reduce the failure mode that drives most of the single-model dissent. Spot-checking the 1500-item sample would convert the flagging rate into a measurable false-positive estimate. Running the same ensemble on items returned by an analogous SPARQL query for physics or computer science would make the cross-domain behavior of the judges directly observable, rather than leaving it to leak through the MathWorld qualitative analysis.

\paragraph{Conclusion.}
An LLM voting ensemble is a practical filter for mathematical concept data imported from Wikidata. The three categories of disagreement identified in this paper (degenerate descriptions, narrow scope bias, editorial-scope mismatches) point to concrete, separable improvements in both the data pipeline and the evaluation methodology.

\begin{credits}
\subsubsection{\ackname} This project began at Dagstuhl Seminar 23401 \emph{Automated Mathematics: Integrating Proofs, Algorithms, and Data}, and the authors gratefully acknowledge the conversations and the collaborative environment there that set Mathswitch in motion. In particular, we are grateful to Andrej Bauer for the encouragement that first prompted us to take this on. We thank Ljup\v{c}o Todorovski for guidance on the design of the experiment and the presentation of results. We thank Tom Wiesing for providing an initial SPARQL query
for retrieving mathematical concepts from Wikidata and thus giving us a working starting point. We also thank the Agda-Unimath editors, particularly Egbert Rijke, Vojt\v{e}ch \v{S}t\v{e}pan\v{c}\'{\i}k, and Fredrik Bakke, for annotating their concepts and making the concept index available.

This material is based upon work supported by the Air Force Office of Scientific Research under award number FA9550-21-1-0024, 
by the Slovenian Research and Innovation Agency (program no.\ P1-0294), and
by the AI For Math Fund, a program of Renaissance Philanthropy.

\end{credits}
%
%
%
\bibliographystyle{splncs04}
\bibliography{bibliography}

\clearpage
\appendix

\section{An example SPARQL query}
\label{app:sparql}
The query can be tested on the Wikidata Query Service\footnote{\url{https://query.wikidata.org/}}. A copyable version is available in the project documentation.\footnote{\url{https://github.com/katjabercic/mathswitch\#query-example-from-the-paper}}

\begin{lstlisting}[
  caption={SPARQL query for one of the property paths. The other path is analogous, with the same optional fields and filters.}]
SELECT DISTINCT
  ?item ?label ?desc ?image ?wp_en ?nlab ?mw ?pw ?eom
  (GROUP_CONCAT(DISTINCT ?alt; separator=", ") AS ?aliases)
WHERE {
  # Property path: 
  # instances of a topic studied by mathematics
  ?item  wdt:P31   ?topic .
  ?topic wdt:P2579 wd:Q395 .
  # External identifiers (optional)
  OPTIONAL { ?item wdt:P4215 ?nlab . }
  OPTIONAL { ?item wdt:P2812 ?mw   . }
  OPTIONAL { ?item wdt:P6781 ?pw   . }
  OPTIONAL { ?item wdt:P7554 ?eom  . }
  # Image and Wikipedia article (optional)
  OPTIONAL { ?item wdt:P18 ?image . }
  OPTIONAL {
    ?wp_en rdf:type schema:Article ;
           schema:isPartOf
             <https://en.wikipedia.org/> ;
           schema:about ?item .
  }
  OPTIONAL { # Aliases
    ?item skos:altLabel ?alt .
    FILTER (lang(?alt) = "en")
  } 
  FILTER NOT EXISTS { # Exclusions
    VALUES ?excludedType {
      wd:Q21199 wd:Q28920044
      wd:Q6256  wd:Q714737
    }
    ?item wdt:P31 ?excludedType .
  }
  FILTER NOT EXISTS { ?item wdt:P31 wd:Q5 . }
  SERVICE wikibase:label {
    bd:serviceParam wikibase:language "en" .
  }
}
GROUP BY ?item ?label ?desc ?image ?wp_en ?nlab ?mw ?pw ?eom
ORDER BY ?item
LIMIT 1000 OFFSET 0
\end{lstlisting}

\pagebreak
\section{Prompt template}
\label{app:prompt}

The following is the full prompt sent to each judge for the Wikidata item \emph{Wiener sausage}.

\begin{verbatim}
You are a categorization judge. Your task is to evaluate whether a 
given concept satisfies a specific predicate.

Be careful with concepts from adjacent domains such as physics, 
computer science, or engineering. A concept should only be 
classified as mathematical if it is primarily mathematical in 
nature. Concepts that merely use mathematics as a tool (e.g. 
quantum mechanics, signal processing) should not be considered 
mathematical concepts. When in doubt, consider whether the 
concept originates from or is primarily studied within mathematics.

You must respond with a structured answer containing:
1. answer: yes or no
2. confidence: a number from 0 to 100 (representing your confidence
   percentage)
3. reasoning: a brief explanation of why you chose that answer

IMPORTANT: Format your response as three lines, exactly like this:
answer: yes
confidence: 85
reasoning: The concept is clearly mathematical because...

---

CONCEPT INFORMATION:
Name: Wiener sausage
Description: Mathematical concept

---

PREDICATE TO EVALUATE:
Is the given concept a mathematical concept, given the name, 
description, keywords, and article text?

---

Please provide your evaluation in the format specified above.
\end{verbatim}

\pagebreak
\section{A selection of examples of LLM judge disagreement}
\label{app:example-disagreements}

\noindent
\textbf{Wiener sausage}

\noindent
DeepSeek-R1 14B's: \emph{yes} (confidence of 95\%). Explanation: \newline
{\footnotesize
The term "Wiener sausage" refers to a concept primarily studied within mathematics, specifically related to stochastic processes and probability theory. It originates from the work of Norbert Wiener and is recognized as a mathematical concept in resources like MathWorld.}

\noindent
Qwen-2.5 14B's: \emph{yes} (confidence of 85\%). Explanation: \newline
{\footnotesize The concept is clearly mathematical because it has a MathWorld ID, indicating it is studied within mathematics.}

\noindent
Gemma-3 12B's: \emph{no} (confidence of 85\%). Explanation: \newline
{\footnotesize The name "Wiener sausage" is a distraction. The MathWorld ID indicates this is a trick entry designed to test users' ability to identify non-mathematical concepts. It's a food item, not a mathematical construct.}

\vspace{0.5cm}
\noindent
\textbf{Turing completeness}

\noindent
DeepSeek-R1 14B's: \emph{yes} (confidence of 75\%). Explanation: \newline
{\footnotesize Turing completeness is primarily studied within theoretical computer science but has strong mathematical foundations, particularly in computability theory. It involves abstract models of computation, making it a mathematical concept.}

\noindent
Qwen-2.5 14B's: \emph{yes} (confidence of 85\%). Explanation: \newline
{\footnotesize Turing completeness is fundamentally based on theoretical computer science but originates from the work of mathematician Alan Turing, making it primarily a mathematical concept.}

\noindent
Gemma-3 12B's: \emph{yes} (confidence of 75\%). Explanation: \newline
{\footnotesize While Turing completeness utilizes mathematical logic and computation, it fundamentally describes a property of computing systems. The concept arises from computer science and theoretical computer science, not primarily from mathematics itself. It's a tool *using* mathematics, rather than a mathematical concept.}

\vspace{0.5cm}
\noindent
\textbf{Boxcars}

\noindent
DeepSeek-R1 14B's: \emph{yes} (confidence 85\%). Explanation: \newline
{\footnotesize The concept relates to dice outcomes, which involves mathematical principles like probability.}

\noindent
Qwen-2.5 14B's: \emph{no} (confidence 30\%). Explanation: \newline
{\footnotesize The term "boxcars" refers to a pair of sixes in dice games, which is more related to gambling or recreational mathematics rather than being a concept primarily studied within mathematics.}

\noindent
Gemma-3 12B's: \emph{no} (confidence 90\%). Explanation: 
{\footnotesize While "boxcars" is a term with a defined probability in dice games, the concept itself originates from and is primarily a feature of games of chance, not mathematics itself. Mathematics is used to analyze the probability, but the core concept is not inherently mathematical.}

\end{document}